\documentclass[fleqn,10pt]{wlscirep}
\usepackage[utf8]{inputenc}
\usepackage[T1]{fontenc}
\usepackage{todonotes}
\usepackage{booktabs}
\usepackage{colortbl}
\usepackage{multirow}
\usepackage{placeins}
\usepackage{amsmath}
\usepackage{amssymb}
\usepackage{amsfonts}
\usepackage{hyperref}
\usepackage{boldline}
\usepackage{cellspace}
\usepackage{color}

\title{fRegGAN with K-space Loss Regularization for Medical Image Translation}
\author[1,*]{Ivo~M.~Baltruschat}
\author[1]{Felix~Kreis}
\author[1]{Alexander~Hoelscher}
\author[1]{Melanie~Dohmen}
\author[1]{Matthias~Lenga}
\affil[1]{Bayer AG, Müllerstr. 178, 13353 Berlin, Germany}
\affil[*]{ivo.baltruschat@bayer.com}
\begin{abstract}
Generative adversarial networks (GANs) have shown remarkable success in generating realistic images and are increasingly used in medical imaging for image-to-image translation tasks. However, GANs tend to suffer from a frequency bias towards low frequencies, which can lead to the removal of important structures in the generated images. To address this issue, we propose a novel frequency-aware image-to-image translation framework based on the supervised RegGAN approach, which we call fRegGAN. The framework employs a K-space loss to regularize the frequency content of the generated images. It incorporates well-known properties of MRI K-space geometry to guide the network training process. By combining our method with the RegGAN approach, we can mitigate the effect of training with misaligned data and frequency bias at the same time. We evaluate our method on the public BraTS dataset and outperform the baseline methods in quantitative and qualitative metrics when synthesizing T2-weighted from T1-weighted MR images. Detailed ablation studies are provided to understand the effect of each modification on the final performance. The proposed method is a step toward improving the performance of image-to-image translation and synthesis in the medical domain.
\end{abstract}
\begin{document}
\flushbottom
\maketitle
\thispagestyle{empty}
\section{Introduction}
\begin{figure}

    \includegraphics[clip, trim=0cm 3.5cm 3.4cm 0.0cm, width=1.00\textwidth]{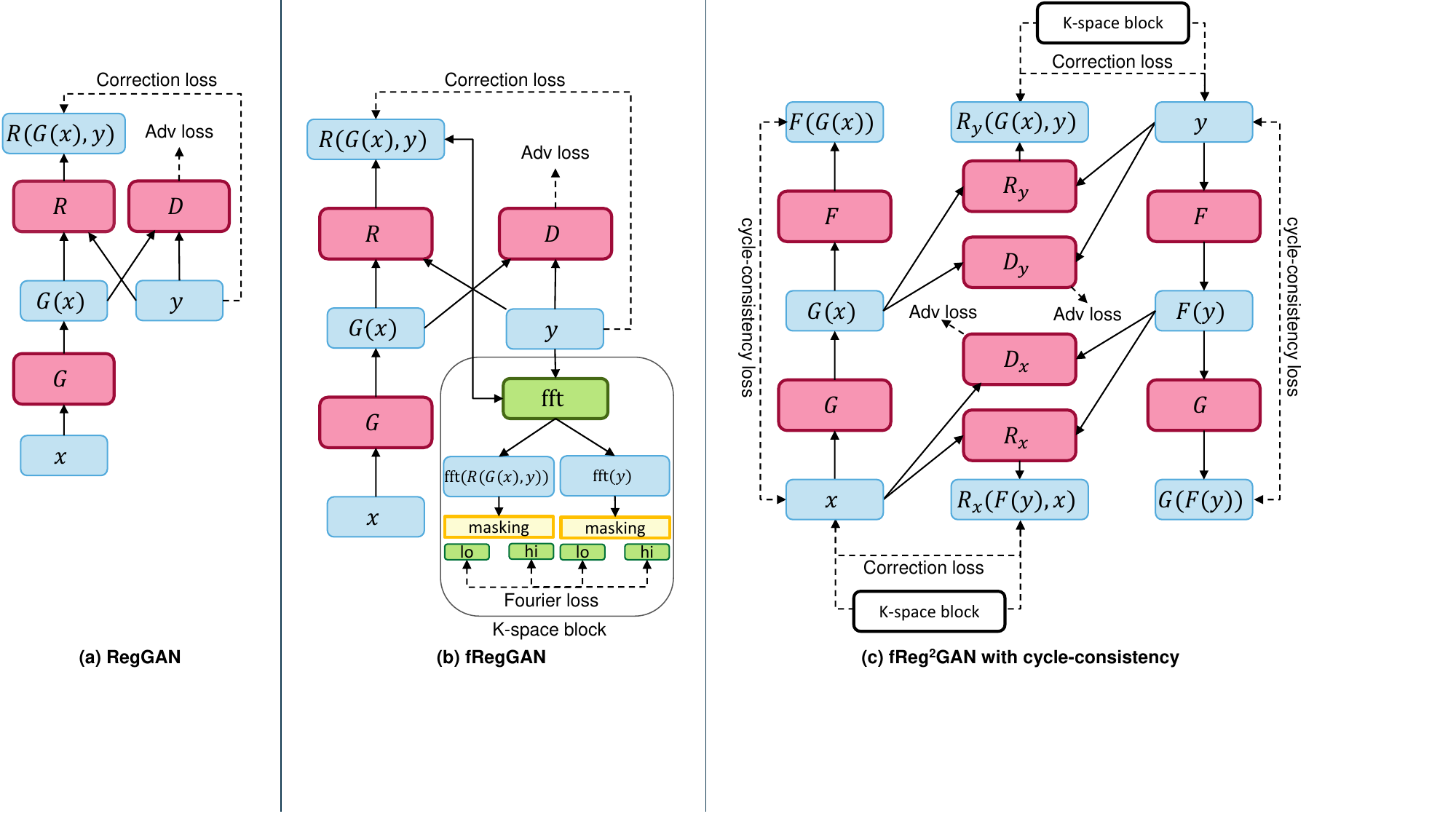}
    
    \caption{Architecture comparison of RegGAN\cite{kongBreakingDilemmaMedical2021} to the proposed methods fRegGAN and fReg$^2$CycleGAN.}\label{fig1}
\end{figure}

Generative adversarial networks (GANs)\cite{goodfellowGenerativeAdversarialNetworks2014} have gained much attention in the last few years due to their ability to generate realistic images. In the medical imaging field, they are often used for image-to-image translation problems, such as mapping magnetic resonance imaging (MRI) to computed tomography images\cite{masperoDoseEvaluationFast2018,yangUnsupervisedMRtoCTSynthesis2020}, T1-weighted to T2-weighted MRI\cite{liuUnifiedConditionalDisentanglement2021,shenMultiDomainImageCompletion2021}, or low- to high dose contrast-enhanced MRIs\cite{hauboldContrastAgentDose2023,pasumarthiGenericDeepLearning2021}. Such approaches could potentially reduce healthcare costs and patient burden while maintaining or even improving the diagnostic value of a modality. While GANs have shown promising results in these tasks, they tend to suffer from a frequency bias towards low frequencies\cite{schwarzFrequencyBiasGenerative2021a}. This is especially problematic for medical image-to-image translation tasks, where preserving the images' high-frequency content (i.e., edges) is crucial. Local errors in the generated images can lead to removing essential structures, such as lesions, which can significantly impact downstream tasks. To address this problem, we propose a novel frequency-aware image-to-image translation framework based on the supervised RegGAN framework\cite{kongBreakingDilemmaMedical2021}. We employ a K-space loss, which can regularize the frequency content of the generated images (see Figure~\ref{fig1}).
Obtaining well-aligned data is often a challenging task in medical image-to-image translation. Common image misalignments, like poor registration of source and target domain samples, can lead to significant degradation in the performance of supervised approaches like Pix2Pix\cite{isolaImagetoImageTranslationConditional2018}, which rely heavily on pixel-wise reconstruction losses. While unsupervised training methods like CycleGAN\cite{zhuUnpairedImagetoImageTranslation2020} can alleviate this issue, they generally tend to have lower performance compared to supervised methods. Huang et al.\cite{kongBreakingDilemmaMedical2021} proposed the supervised RegGAN medical image-to-image translation approach. In the RegGAN framework, the misaligned target images are considered noisy labels, and the generator is trained with an additional registration network estimating a displacement vector field to fit the misaligned noise distribution adaptively.
We are not the first to investigate the use of the frequency domain for improving the performance of GANs, but our work is the first to apply the frequency domain to the medical image-to-image translation task and to combine it with the RegGAN.
In the field of image processing and generation, several recent studies \cite{yangFDAFourierDomain2020,jiangFocalFrequencyLoss2021,yangFreGANExploitingFrequency2022} have explored the use of the frequency domain for improving the performance of various models.
The closest to our idea is in the work of Cai et al. \cite{caiFrequencyDomainImage2021}, where a frequency domain image translation framework was proposed for image-to-image translation. The framework exploits frequency information by adding multiple losses based on the frequency-transformed images to the optimization problem of the generator. They showed natural images that the proposed method could improve the performance of image-to-image translation and synthesis. An open question is if the proposed method can be combined with cycle consistency and applied to medical images.
We summarize our contributions as follows:
\begin{enumerate}
    \item We further investigate the RegGAN approach by Huang et al. \cite{kongBreakingDilemmaMedical2021} by generally applicable modifications to the architecture and training procedure. We extend their CycleGAN approach by adding a second registration network, which we call Reg$^2$CycleGAN.
    \item We incorporate constraints in the frequency domain (K-space) to regularize and guide the network training process. This regularization technique is motivated by MRI acquisition and reconstruction methods which leverage the distribution of image feature information in K-space to reduce noise and/or accelerate image acquisition.
    \item Our proposed method is evaluated on the publicly available BraTS dataset. The results show that our method outperforms the baseline (i.e., RegGAN and CycleGAN) methods in quantitative and qualitative metrics. We also provide detailed ablation studies to understand the effect of each modification on the final performance. Our proposed method is a step towards improving the performance of image-to-image translation and synthesis in the medical domain to bring GAN-based methods closer to clinical application.
\end{enumerate}
\section{Methods}
\label{methods}
This section begins with a brief overview of how vanilla GANs\cite{goodfellowGenerativeAdversarialNetworks2014} are formulated and then summarizes later improvements like the LSGAN\cite{maoLeastSquaresGenerative2017} and RegGAN\cite{caiFrequencyDomainImage2021}. Afterward, we describe our new K-space Loss regularization and its integration into the RegGAN and the CycleGAN.
GANs are generative models composed of two networks, a generator $G$ and a discriminator $D$. The generator learns a mapping $G: X \rightarrow Y$  from source domain $X$ to target domain $Y$ such that $\widehat{y} = G(x)$ is close to $y$ w.r.t. a specific metric. The discriminator $D$ is a binary classifier trained to distinguish between real $y$ and generated $\widehat{y}$. Here, $x$ and $y$ denote sets of paired images $\{(x_i, y_i)\}_{i=1}^N$ where $x_i$ is an image in the source domain $X$ and $y_i$ is the corresponding image in the target domain $Y$. Following the notation from Kong et al.\cite{kongBreakingDilemmaMedical2021}, the vanilla GAN optimization problem is:
\begin{equation}
    \label{eq:vanilla_gan}
    \min_{G} \max_{D} \mathcal{L}_{GAN}(D,G) = \mathbb{E}_{y} \log D(y)  + \mathbb{E}_{x}\log \left( 1 - D(G(x)) \right)
\end{equation}
Since optimizing $G$ and $D$ simultaneously is not possible, the optimization problem is solved by alternating between $D$ and $G$\cite{goodfellowGenerativeAdversarialNetworks2014}. The vanilla GAN training procedure has stability issues, which motivated the proposition of several modifications. The most relevant for our experiments is the Least Squares GAN (LSGAN)\cite{maoLeastSquaresGenerative2017}, which replaces the binary cross entropy loss with the mean squared error loss, resulting in the optimization task:
\begin{equation}
    \begin{aligned}
         \min_{D} \mathcal{L}_{D,LSGAN}(D,G) &= \mathbb{E}_{y} \left( D(y) - 1 \right)^2  + \mathbb{E}_{x} \left( D(G(x)) \right)^2  \\
         \min_{G} \mathcal{L}_{G,LSGAN}(D,G) &= \mathbb{E}_{x} \left( D(G(x)) - 1 \right)^2
    \end{aligned}
\end{equation}
In the work of Kong et al.\cite{kongBreakingDilemmaMedical2021}, the robustness of Pix2Pix\cite{isolaImagetoImageTranslationConditional2018}, which uses an L1-loss for supervision, was analyzed in the presence of misaligned source and target domain image data. Based on their experiments on the publicly available BraTS benchmark, it is concluded that the L1-loss reconstruction term only works with well-aligned images, which is often not the case in medical imaging. Therefore, the RegGAN method proposes to replace the L1-loss with the following correction loss:
\begin{equation}
    \min_{G,R} \mathcal{L}_{corr}(G,R) = \mathbb{E}_{x, y} \left| y - G(x) \circ R(G(x), y) \right|
\end{equation}
Here $R(G(x), y)$ denotes a deformation vector field (DVF) which is estimated based on $G(x), y$ using VoxelMorph\cite{balakrishnanVoxelMorphLearningFramework2019} and $\circ$ is the resample operator.
Adding the smoothness constraint $\mathcal{L}_{smooth}(R) = \mathbb{E}_{x, y} \left| \nabla R(G(x), y) \right|^2$ for the DVF, the final RegGAN optimization problem is then formulated as follows:
\begin{equation}
    \label{eq:reggan}
    \begin{aligned}
    \min_{D} \mathcal{L}_{D,reg}(D,G) &= \mathcal{L}_{D,LSGAN}(D,G) = \mathbb{E}_{y} \left( D(y) - 1 \right)^2  + \mathbb{E}_{x} \left( D(G(x)) \right)^2  \\
    \min_{G,R} \mathcal{L}_{G,reg}(D,G,R) &= \mathcal{L}_{G,LSGAN}(D,G) + \lambda_{1} \mathcal{L}_{corr}(G,R) + \lambda_{2} \mathcal{L}_{smooth}(R)
    \end{aligned}
\end{equation}
For unsupervised image-to-image translation, Zhu et al.\cite{zhuUnpairedImagetoImageTranslation2020} proposed to use two generators $F, G$, and two discriminators $D_{X}, D_{Y}$, and to add a cycle consistency loss $\mathcal{L}_{cyc}(G,F) = \mathbb{E}_{x}  \left| F(G(x)) - y \right|   + \mathbb{E}_{y}  \left| G(F(y)) - x \right|  $. In additional experiments, they showed that an identity loss $\mathcal{L}_{id}(G, F) = \mathbb{E}_{y}  \left| G(y) - y \right|   + \mathbb{E}_{x}  \left| F(x) - x \right|  $ preserves content information. All four models are trained jointly, and the optimization problem for the generators is formulated as follows:
\begin{equation}
    \label{eq:cyclegan}
    \min_{G,F} \mathcal{L}_{G,CycGAN}(D_{X}, D_{Y}, G, F) = \mathcal{L}_{G,LSGAN}(D_{Y},G)
    + \mathcal{L}_{G,LSGAN}(D_{X},F)
    + \lambda_{3} \mathcal{L}_{cyc}(G,F)
    + \lambda_{4} \mathcal{L}_{id}(G,F)
\end{equation}
Kong et al.\cite{kongBreakingDilemmaMedical2021} also proposed combining cycle consistency with their registration loss
\begin{equation}
    \label{eq:regcyclegan}
    \min_{G, F, R} \mathcal{L}_{G,RegCycGAN}(D_{X}, D_{Y}, G, F, R) =
    \mathcal{L}_{G,reg}(D_{Y},G, R)
    + \mathcal{L}_{G,LSGAN}(D_{X},F)
    + \lambda_{3} \mathcal{L}_{cyc}(G,F)
    + \lambda_{4} \mathcal{L}_{id}(G,F)
\end{equation}
which turns the unsupervised training task into a supervised one. This has the beneficial effect of reducing the number of solutions for the generators and improving the overall quality.
\subsection{Reg$^2$CycleGAN}
\label{sec:reg2cyclegan}
To further improve the performance of the registration and supervision of the generators in equation~(\ref{eq:regcyclegan}), we propose an extension to the existing method. As shown in Figure~\ref{fig1}, we suggest expanding the registration loss to generator $F$. Both $F$ and $G$ suffer from the same data misalignment, which can negatively impact their performance when training with a supervised loss. By modifying the optimization problem described above, we can ensure that both generators are subjected to the same level of registration and supervision. This can help improve the overall stability and reliability of the generated images, as well as ensure that both generators are able to learn from each other's mistakes and improve their performance over time. Here, cycle consistency links the optimization of both generators, meaning that improvements to one will likely result in improvements to the other. So, not only will $F$ benefit from the improved registration loss, but $G$ should also see a positive impact. We can formulate the optimization problem as:
\begin{equation}
    \label{eq:regregcyclegan}
    \begin{split}
    \min_{G, F, R_{X}, R_{Y}} \mathcal{L}_{G, Reg^2CycGan}(D_{X}, D_{Y}, G, F, R_{X}, R_{Y}) =
    \mathcal{L}_{G,reg}(D_{Y}, G, R_{Y}) + \\ \mathcal{L}_{G,reg}(D_{X}, F, R_{X})
    + \lambda_{3} \mathcal{L}_{cyc}(G,F) + \lambda_{4} \mathcal{L}_{idy}(G,F)
\end{split}
\end{equation}
\subsection{Frequency regularization}
\begin{figure}
  \begin{center}
    \includegraphics[width=0.9\textwidth]{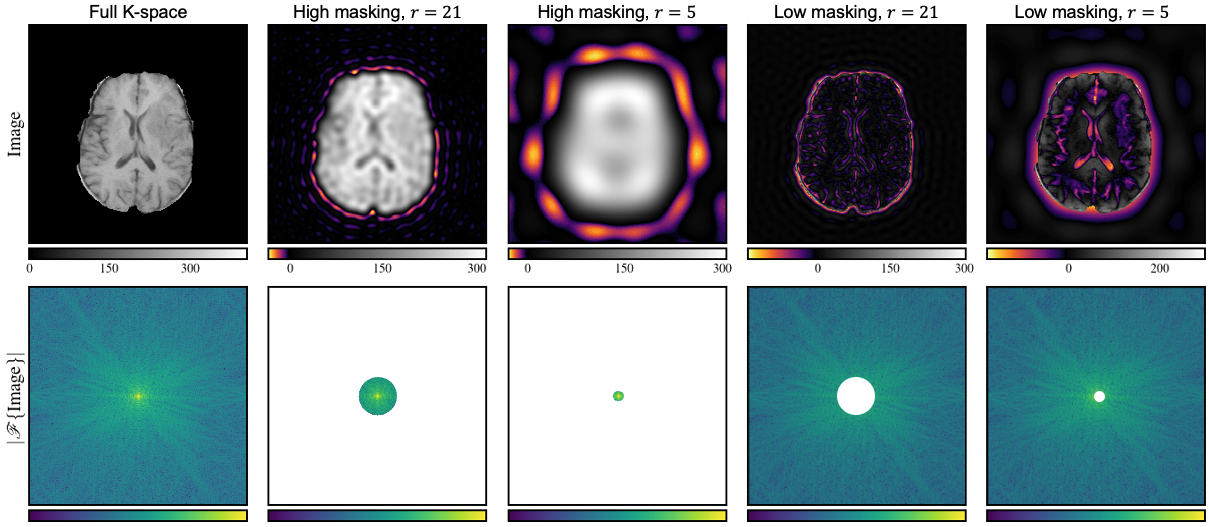}
  \end{center}
  
  \caption{The effects of K-space masking with a variable radius.  Bottom row: (zero-masked) K-spaces. Top row: images reconstructed from (zero-masked) K-spaces.
  From left to right: full K-space, zero-masking of K-space components based on the radius, i.e., setting $r=21$,$w_{freq}=1$; $r=5$, $w_{freq}=1$; $r=21$,$w_{freq}=0$, and $r=5$,$w_{freq}=0$.} \label{fig4}
\end{figure}

\begin{figure}
  \begin{center}
    \includegraphics[width=0.9\textwidth]{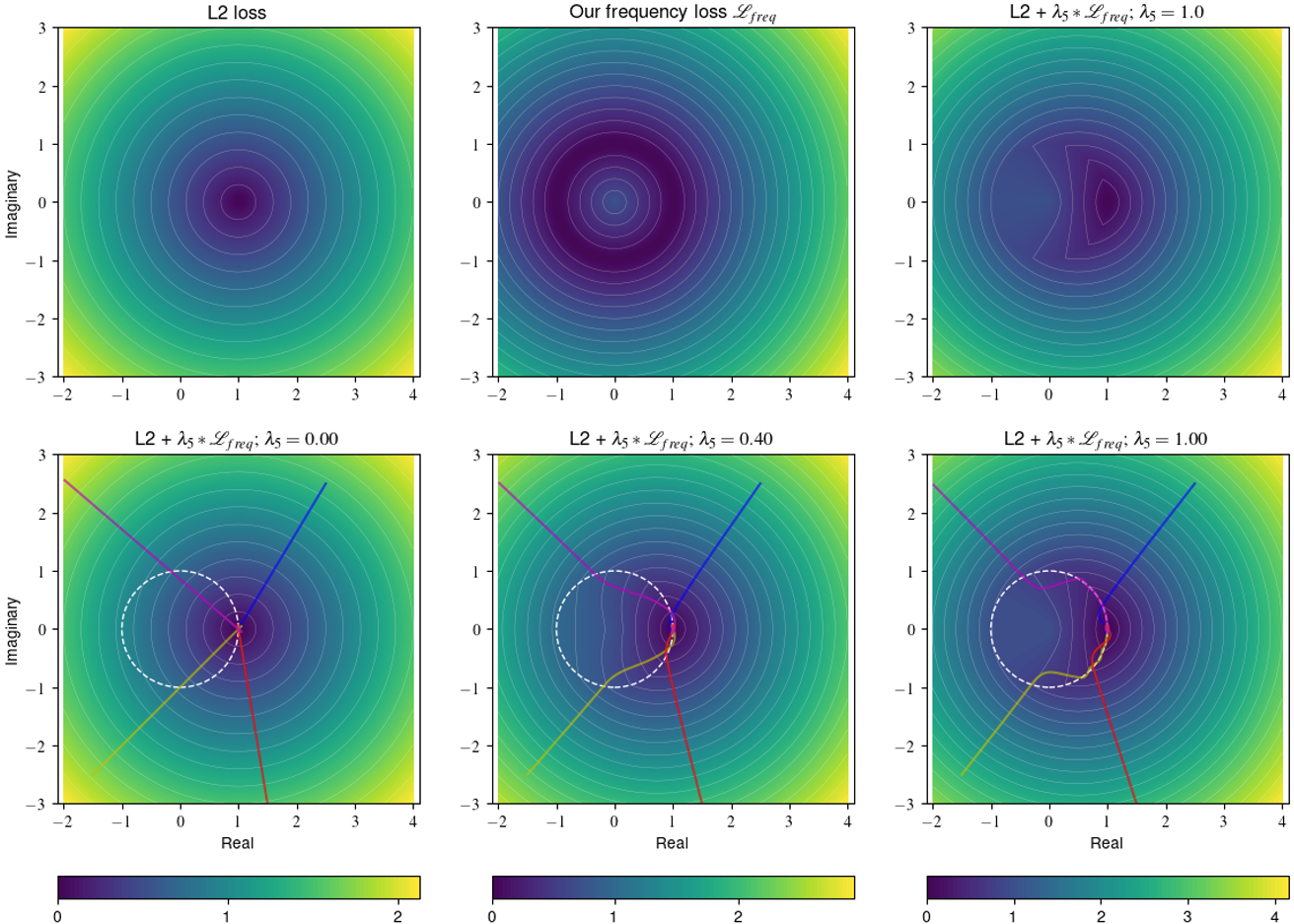}
  \end{center}
  
  \caption{Loss landscape and the impact on gradient descent. Top row: loss landscape for L2, our proposed frequency loss, and the combination of both $\text{L}_2 + \lambda_5 \mathcal{L}_\textit{freq}$. Bottom row: gradient descent optimization starting from four different points. For all images, we used the target point $p = 1 + 0i$.  The white circle indicates the unit sphere $S_1 := \{z : \Vert z \Vert_2 = 1\}$, i.e. the optimal solutions based on our frequency loss, and the star marks the minimum for the L2 loss. $\lambda_5$ is set to $0$, $0.4$, and $1$. } \label{fig5}
\end{figure}

The proposed frequency regularization is motivated by MRI, where the raw signal is a frequency domain signal that is organized in K-space. An inverse Fourier transform of the K-space yields the final image.
The information in K-space is represented by complex-valued base coefficients that result from a waveform decomposition of the reconstructed image. Coefficients near the center of K-space correspond to low-frequency components that capture a large proportion of the image's contrast and texture information. In contrast, coefficients near the outer boundary of K-space correspond to high-frequency features of the image. In MRI pulse sequence development, this knowledge about information content in different regions of K-space is crucial. It leads to a preference for acquiring central K-space regions when acquisition time is limited. Another essential property of the waveform base functions that further motivates the proposed regularization is their non-locality, i.e., locally changing image voxel values globally impact the K-space coefficients, and local changes to K-space coefficients globally affect voxel values.
We define the frequency loss $\mathcal{L}_{freq}$ as the mean L1 distance between the magnitude of the discrete Fourier transform $\mathcal{F}$ of the generated image $G(x)$ and the magnitude of the discrete Fourier transform of the target image $y$. In addition, we use a binary mask $M$, where the circular region around the origin with radius $r$ is set to 1, its inverse $\overline{M}$, and a weighting factor $w_{freq}$. In our experiments below we consider the three scenarios $w_{freq} \in \{f_\text{low}=1,f_\text{hi}=0, f_\text{all}=0.5\}$. The frequency loss is then defined as
\begin{equation}
    \label{eq:frequency_loss}
    \begin{aligned}
             \mathcal{L}_{freq}(G) &=  w_{freq} \, \mathbb{E}_{x}\left\| \left| \mathcal{F}\left\{ G(x) \right\} \right| \odot M  - \left| \mathcal{F}\left\{ y \right\} \right| \odot M \right\|_{1}
    + (1-w_{freq}) \, \mathbb{E}_{x} \left\| \left| \mathcal{F}\left\{ G(x) \right\} \right| \odot \overline{M} - \left| \mathcal{F}\left\{ y \right\} \right|\odot \overline{M} \right\|_{1} \\
    &=   \mathbb{E}_{x}\left\| \left| \mathcal{F}\left\{ G(x) \right\} \right| \odot K   - \left| \mathcal{F}\left\{ y \right\} \right| \odot K \right\|_{1}
    \end{aligned}
\end{equation}
where $\odot$ denotes the Hadamard product, $\left|\mathcal{F}\left\{ \cdot \right\} \right|$ the element-wise absolute value of the complex tensor $\mathcal{F}\left\{ \cdot \right\}$ and $K:=  w_{freq}  M  + (1-w_{freq}) \overline{M}$ our selected window function. The given choice of $K$ implies a distinct penalization of deviations of the low and high-frequency components. The parameter $r$ controls the separation boundary, whereas the weighting factor $w_{freq}$ balances the impact of those two classes.  Clearly, other choices of $K$ are possible, e.g. Gaussian kernel or a Hann window. To provide more intuition about our choice of window function,we show in Figure~\ref{fig4} the effect of K-space masking with a variable radius. We can see that the masking of the high-frequency components leads to a blurry image. On the other hand, the masking of the low-frequency components leads to a loss of contrast and texture information.
Our frequency loss \eqref{eq:frequency_loss} has some beneficial properties compared to losses in the image domain. First, $\mathcal{L}_{freq}$ is translation invariant because we only consider the magnitude, not the phase. Secondly, it is more sensitive to blur than to mild noise because blurry solutions have a higher impact on the high-frequency components.
In order to further study the effect of the proposed frequency regularization, we compare a simple L2 loss  with our frequency loss and the combination of both. Please note that the L2 loss is equivalent in image and frequency domain due to Plancherel's theorem.
In the top row of Figure~\ref{fig5}, we calculated the loss value for a complex target point $p= 1 + 0i$ and various predictions. On the x-axis, we show the real part, and on the y-axis, we show the imaginary part. The color shows the loss value. The L2 loss has circular level sets and increases as we move away from the unique minimum $p$. In contrast, the frequency loss has multiple minimums. All points which have the same magnitude as the target point, i.e. points located on the sphere $S:=\{z : \Vert z \Vert_2 = \Vert p \Vert_2 \}$, are minimizers.
In the bottom row of Figure~\ref{fig5}, we show a simple gradient descent with momentum optimization for four different starting points. The white circle corresponds to $S$, i.e the optimal solutions based on our frequency loss, and the star marks the minimum for the L2 loss. We combine the L2 and the frequency loss using different different weighting factors $\lambda_5 \in \{0, 0.4, 1\}$ in $\text{L}_2 + \lambda_5 \mathcal{L}_\textit{freq}$. For $\lambda_5=0$, the frequency loss is not considered, and the optimization converges to the L2 minimum. For $\lambda_5=0.4$ and $\lambda_5=1$ the frequency loss is added, we can observe that our new loss acts as a regularizer. The optimization converges to the minimum of the frequency loss and follows the manifold $S$ to the optimal solution. The higher $\lambda_5$, the more the optimization is drawn to $S$ and to take gradient steps along this surface.
Like the other supervised losses, the frequency loss can not be used if the data pair $(x,y)$ is not well aligned (e.g., rotation). Therefore, we propose to replace the generated image $G(x)$ with its registered version $G(x) \circ R(G(x),y)$ in this case. In our experiments, we add $\lambda_{5} \mathcal{L}_{freq}(G)$ and $\lambda_{5} (\mathcal{L}_{freq}(G) + \mathcal{L}_{freq}(F))$ to the total loss of the GAN and CycleGAN training, which yields the final fRegGAN and fReg$^2$CycleGAN architectures as depicted in Figure \ref{fig1}.
\FloatBarrier
\section{Experiments and Results}
We perform two sets of experiments. First, we set a baseline and solely investigate our changes to the Reg$^2$CycleGAN. Secondly, we explore the combination of the RegGAN and Reg$^2$CycleGAN with our K-space loss (see equation~(\ref{eq:frequency_loss})). All our experiments are trained on the public BraTS 2021 dataset\cite{baidRSNAASNRMICCAIBraTS20212021,menzeMultimodalBrainTumor2015,bakasAdvancingCancerGenome2017}. BraTS has 1251 MRI scans of patients with brain tumors. The dataset contains four different modalities, but we only use the T1w and T2w modalities to train our models. All MRI scans are available as NIfTI files and were acquired with other clinical protocols and various scanners from multiple data-contributing institutions. In addition, all volumes are pre-processed, i.e., co-registered to the same anatomical template, interpolated to the same resolution ($1\text{ mm}^3$), and skull-stripped. Hence, the dataset is well aligned, and each volume has a size of $240 \times 240 \times 155$ voxels.
To train and evaluate our models, we randomly split the dataset on patient-level into a training, validation, and testing set with 1000, 51, and 200 volumes, respectively. We further pre-process the dataset by normalizing each volume to the range $[0, 1]$ based on the $0.5$ and $99.5$ percentile of the volume. Afterward, we sliced the 3d volumes in axial direction into 2d images and removed all slices without information. This results in 139221, 6891, and 27956 2d slices for training, validation, and testing, respectively.
Based on the survey papers\cite{liu3DBrainHeart2022,xieCrossModalityNeuroimageSynthesis2022,aliRoleGenerativeAdversarial2022}, we selected to most used metrics for evaluating our models: peak signal-to-noise ratio (PSNR), structural similarity index (SSIM), and multi-scale structural similarity index (MS-SSIM).
The SSIM is a measure of the similarity between two images. It is based on the mean and variance of the local luminance and contrast of the images. The MS-SSIM is a multi-scale version of the SSIM. It is defined as the geometric mean of the SSIM at different scales. The SSIM and MS-SSIM are in the range $[0, 1]$ and a higher value indicates a better synthetization. Both metrics are implemented using the TorchMetrics~0.10.3 library\cite{detlefsenTorchMetricsMeasuringReproducibility2022}. 
\subsection*{Implementation:}
\label{sec:implementation}
We used a fixed training setup for all experiments to ensure a fair comparison. All models are trained with online data augmentation. Here, we use random rotation between $-10$ and $10$ degrees, random translation between $-26$ and $26$ pixels for x and y direction, and random scaling between $0.9$ and $1.1$. In a second step, the input images and the target images are artificially misaligned by applying the same random transformations as the online data augmentation, but only to the target images. Similar to Kong et al.\cite{kongBreakingDilemmaMedical2021}, we use the term noise for this misalignment in the following.
As optimizer, we use Adam\cite{kingmaAdamMethodStochastic2017} with a learning rate of $0.0001$, $\beta_1$ of $0.5$, $\beta_2$ of $0.999$ and no weight decay. We train all models for $1 \times 10^6$ iterations with a batch size of $4$. For the total loss, we used $\lambda_{1}=20$, $\lambda_2=10$, $\lambda_3=10$, $\lambda_4=1$. Empirically, we found for our K-space loss that $r=21$, $\lambda_5=1$ for $f_{hi}$ and $\lambda_5=0.1$ for $f_{low}$ and $f_{all}$ work well.
To implement our models, we use PyTorch~1.12.1\cite{paszkePyTorchImperativeStyle2019} and the PyTorch Lightning~1.8.1\cite{Falcon_PyTorch_Lightning_2019} libraries. All experiments are preformed on a single NVIDIA Tesla A100 GPU with 40 GB of memory. Our code is publicly available at \url{https://github.com/Bayer-Group/fRegGAN}.
\subsection*{Registration with GAN and CycleGAN}
Table~\ref{tab1} shows in the right block the results of our first set of experiments. Our baseline (i.e., without artificial misalignment) results show for all three metrics PSNR, SSIM, and MS-SSIM that the supervised GAN is performing better than the unsupervised CycleGAN (i.e., 26.0 vs. 24.1, 0.91 vs. 0.89, and 0.94 vs. 0.91). This is expected since CycleGAN is trained without any supervision. However, adding noise (i.e., artificial misalignment) to the training, we can observe that the situation is vice-versa (i.e., 21.9 vs. 23.8, 0.83 vs. 0.88, and 0.84 vs. 0.90). Both observations are in line with other results in the literature\cite{kongBreakingDilemmaMedical2021}. As described in Section~\ref{methods}, adding a registration loss to the CycleGan (i.e., RegCycleGAN) makes it also supervised. Now, the supervised RegCycleGAN is performing better than the supervised GAN (i.e., 27.4 vs. 26.0, 0.92 vs. 0.91, and 0.95 vs. 0.94). The cycle consistency combined with supervised learning is a strong regularization term that helps to learn the mapping between the input and the target image. Adding a second registration loss (i.e., Reg$^2$CycleGAN) to the RegCycleGAN improves the results of the generator $F$ but does not help to further improve our main generator $G$. The second registration loss has only an indirect connection to the first generator $G$ by the cycle consistency, which is not enough to improve the results. 
\begin{table}\centering
    \caption{Quantitative results for training the GAN and CycleGAN with (w/) and without (w/o) the following features: Registration and noise (i.e., artificial misalignment, see Section \ref{sec:reg2cyclegan}). For example, a model trained with registration and noise is marked as ''w/ n, r". We show for the CycleGAN models the results for both generators (i.e., G and F). The bold text emphasizes the best overall result for each metric. As a reference, we show the results from Kong et al.\cite{kongBreakingDilemmaMedical2021}. $^\dagger$ model is a GAN and not a cGAN like in Pix2Pix. *model was trained with less misalignment}\label{tab1}
    \begin{tabular}{clc|cc | cc}\toprule
                            &  &             & \multicolumn{2}{c}{\bf{CycleGAN}} & & \\
                            &  & \bf{GAN}   & \bf{$G$} & \bf{$F$} & \bf{Pix2Pix}\cite{kongBreakingDilemmaMedical2021} & \bf{CycleGAN}\cite{kongBreakingDilemmaMedical2021} \\ \hlineB{1.25} \rowcolor[gray]{.95}
                            & baseline  & 26.0  & 24.1      & 23.9  & 25.6 & 23.9 \\
                            & w/ n     & 21.9  & 23.8      & 22.9  & 15.0 & 23.7 \\ \rowcolor[gray]{.95}
                            & w/ n, r       & 26.8  & \bf{27.4} & 23.2  & 25.2$^\dagger$ & 23.8* \\
    \multirow{-4}{*}{PSNR}  & w/ n, r$^2$   & -     & 27.3      & 27.3  & - & - \\ \hlineB{1.25} \rowcolor[gray]{.95}
                            & baseline  & 0.91  & 0.89      & 0.89  & 0.85 & 0.83 \\
                            & w/ n     & 0.83  & 0.88      & 0.87  & 0.74 & 0.83\\ \rowcolor[gray]{.95}
                            & w/ n, r       & 0.91  & \bf{0.92} & 0.89  & 0.85$^\dagger$ & 0.85*\\
    \multirow{-4}{*}{SSIM}  & w/ n, r$^2$   & -     & \bf{0.92} & 0.90  & - & - \\ \hlineB{1.25} \rowcolor[gray]{.95}
                            & baseline  & 0.94  & 0.91      & 0.93  & - & - \\
                            & w/ n     & 0.84  & 0.90      & 0.92  & - & - \\ \rowcolor[gray]{.95}
                            & w/ n, r       & 0.94  & \bf{0.95} & 0.92  & - & - \\
    \multirow{-4}{*}{MS-SSIM} & w/ n, r$^2$ & -     & \bf{0.95} & 0.96  & - & - \\ \bottomrule
    \end{tabular}
\end{table}

\subsection*{K-space loss}
\begin{figure}
    
    \includegraphics[clip, trim=0cm 0cm 0cm 0cm, width=1.00\textwidth]{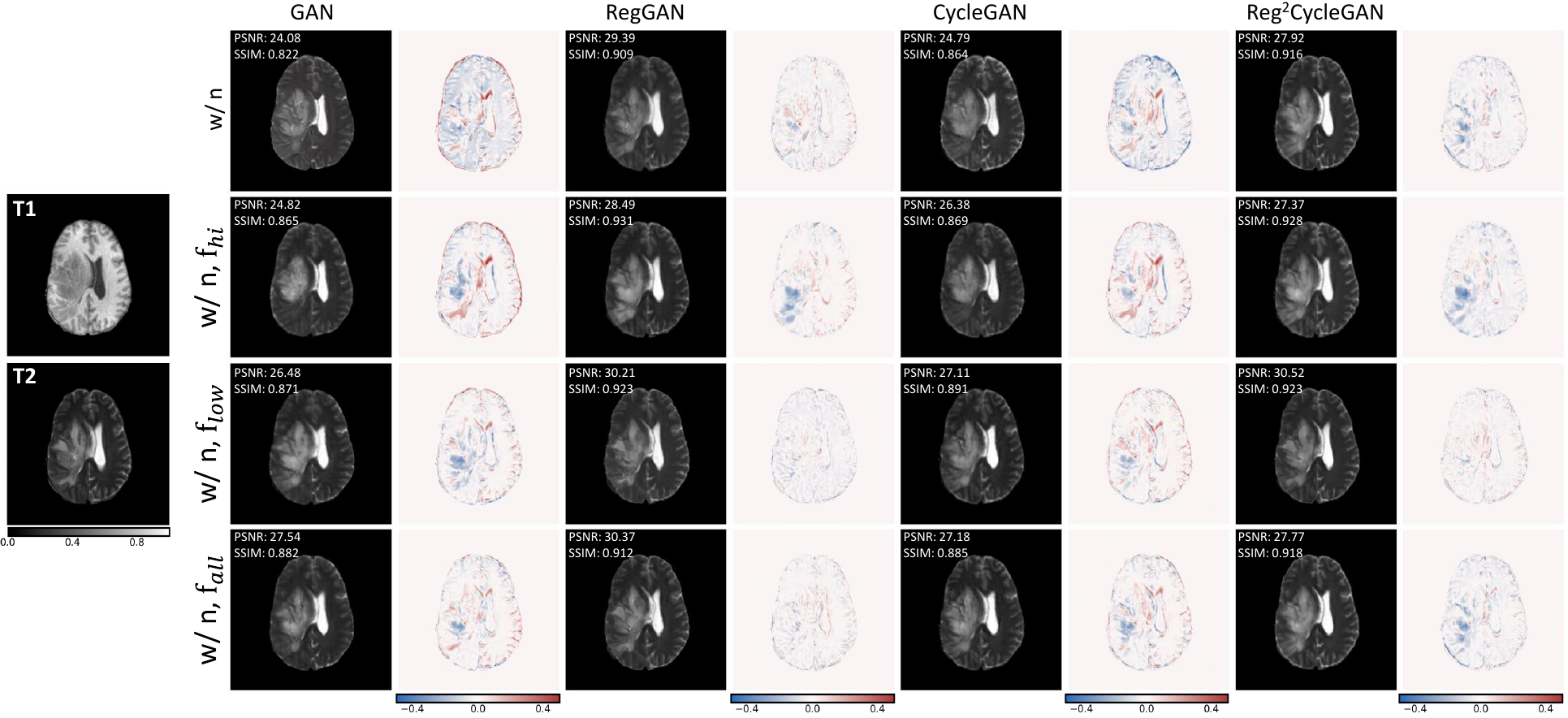}
    \caption{Qualitative results for training our models with K-space loss. The first column shows the T1w input image and the T2w ground truth. Column by column, we show the prediction and the residual to ground truth for each model. Row by row, we show different training setups for each of these models.} \label{fig3}
\end{figure}

We defined three different settings of the K-space loss for our experiments. First, we only select the low frequencies with $w_{freq} = f_\text{low} = 1$. Secondly, the high frequencies are selected by $w_{freq} = f_\text{hi} = 0$. Finally, we also test the effect if all frequencies are used $w_{freq} = f_\text{all}= 0.5$. All three settings are tested for both GAN and CycleGAN models with and without registration loss. The results are shown in
Table~\ref{tab2}. For all four training modes (i.e., GAN with noise, GAN with noise and reg, CycleGAN with noise, and CycleGAN with noise and reg), we observe that adding a frequency loss improves the quantitative metrics. The only exception is the CycleGAN model with $f_\text{hi}$. Here, the PSNR is slightly worse than the baseline (i.e., 23.6 vs 23.8). However, the SSIM and MS-SSIM are the same as the baseline with 0.88 and 0.90, respectively. The best results are achieved by the GAN model with $f_\text{low}$ and the registration loss. The PSNR is 28.8, the SSIM is 0.92, and the MS-SSIM is 0.97. The results show that the frequency loss is useful for improving the quantitative metrics. In Figure~\ref{fig3}, we show qualitative examples of the generated images. Here, we see that the models trained with $f_\text{low}$ and registration can better synthesize the highlighted region.
\begin{table}\centering
    \caption{Quantitative results for training GAN and CycleGAN with K-space loss. We trained the models in two modes: with (w/) noise (i.e., artificial misalignment, see Section~\ref{sec:implementation}) and w/ noise and registration. For each mode, we investigated the contribution of training with different K-space losses: $f_{hi}$, $f_{low}$, $f_{all}$. The bold text emphasizes the best overall result for each metric.}\label{tab2}
    \begin{tabular}{clc|c|c|c}\toprule
           & & \multicolumn{2}{c}{\bf{GAN}} & \multicolumn{2}{c}{\bf{CycleGAN}} \\
           & & w/ n &  w/ n, r & w/ n &  w/ n, r$^2$ \\ \hlineB{1.25} \rowcolor[gray]{.95}
                            & baseline & 21.9 & 26.8 & 23.8 & 27.3 \\
                            & $f_{hi}$ & 22.7 & 26.9 & 23.6 & 27.4\\\rowcolor[gray]{.95}
                            & $f_{low}$ & 24.4 & \bf{28.8} & 25.2 & 28.7\\
    \multirow{-4}{*}{PSNR}  & $f_{all}$ & 24.6 & 28.4 & 25.2 & 28.7\\ \hlineB{1.25} \rowcolor[gray]{.95}
                            & baseline & 0.83 & 0.91 & 0.88 & 0.92 \\
                            & $f_{hi}$ & 0.86 & \bf{0.93} & 0.88 & \bf{0.93} \\\rowcolor[gray]{.95}
                            & $f_{low}$ & 0.88 & 0.92 & 0.89 & 0.92 \\
    \multirow{-4}{*}{SSIM}  & $f_{all}$ & 0.88 & 0.91 & 0.89 & 0.92 \\ \hlineB{1.25} \rowcolor[gray]{.95}
                            & baseline & 0.84 & 0.94 & 0.90 & 0.95 \\
                            & $f_{hi}$ & 0.87 & 0.94 & 0.90 & 0.95 \\\rowcolor[gray]{.95}
                            & $f_{low}$ & 0.91 & \bf{0.97} & 0.93 & 0.96 \\
    \multirow{-4}{*}{MS-SSIM}& $f_{all}$ & 0.92 & 0.96 & 0.93 & 0.96\\ \bottomrule
    \end{tabular}
\end{table}
\section{Conclusion}
We have proposed a novel image-to-image translation approach in the medical imaging field called fRegGAN. Our method extends the RegGAN approach by incorporating a Fourier-space based loss to regularize and guide the network training process. Our extensive experiments on the BraTS 2021 dataset show that our proposed method outperforms the baseline methods in terms of both quantitative metrics and qualitative visual inspection. The frequency regularization loss helps the optimization process find solutions that preserve more details and are closer to the ground truth.
However, in our evaluation, we found that comparison based on quantitative metrics with reference images has some limitations. Skull stripping (to anonymize patients) resulted in some artificial artifacts between input and output volumes (i.e., T1w and T2w). Our results prove that incorporating Fourier-space information can improve image-to-image translation's performance and visual quality in the medical domain.
Our findings suggest that fRegGAN could have broader implications for the medical imaging field. For example, frequency regularization, an essential aspect of our proposed method, might be beneficially applied to image synthesis tasks using diffusion models. The improvements observed in image-to-image translation tasks could potentially translate to enhancements in the quality and accuracy of synthesized images, particularly in complex tasks such as those involving multiple modalities.
Moreover, the idea of k-space to k-space synthesis, although not directly addressed in our study, could serve as a promising future research direction. Such an approach might further aid clinical decision-making, diagnostics, and treatment planning by providing more accurate and realistic synthetic medical images.
\FloatBarrier
\bibliography{FreqLossPaper_short}

\begin{thebibliography}{10}
\urlstyle{rm}
\expandafter\ifx\csname url\endcsname\relax
  \def\url#1{\texttt{#1}}\fi
\expandafter\ifx\csname urlprefix\endcsname\relax\def\urlprefix{URL }\fi
\expandafter\ifx\csname doiprefix\endcsname\relax\def\doiprefix{DOI: }\fi
\providecommand{\bibinfo}[2]{#2}
\providecommand{\eprint}[2][]{\url{#2}}

\bibitem{goodfellowGenerativeAdversarialNetworks2014}
\bibinfo{author}{Goodfellow, I.~J.}, \bibinfo{author}{{Pouget-Abadie}, J.},
  \bibinfo{author}{Mirza, M.} \& \bibinfo{author}{et~al.}
\newblock \bibinfo{journal}{\bibinfo{title}{Generative {{Adversarial
  Networks}}}}.
\newblock {\emph{\JournalTitle{Communications of the ACM}}}
  \textbf{\bibinfo{volume}{63}}, \bibinfo{pages}{139--144},
  \doiprefix\url{10.1145/3422622} (\bibinfo{year}{2020}).

\bibitem{masperoDoseEvaluationFast2018}
\bibinfo{author}{Maspero, M.}, \bibinfo{author}{Savenije, M. H.~F.},
  \bibinfo{author}{Dinkla, A.~M.} \& \bibinfo{author}{et~al.}
\newblock \bibinfo{journal}{\bibinfo{title}{Dose evaluation of fast
  synthetic-{{CT}} generation using a generative adversarial network for
  general pelvis {{MR-only}} radiotherapy}}.
\newblock {\emph{\JournalTitle{Physics in Medicine and Biology}}}
  \textbf{\bibinfo{volume}{63}}, \bibinfo{pages}{185001},
  \doiprefix\url{10.1088/1361-6560/aada6d} (\bibinfo{year}{2018}).

\bibitem{yangUnsupervisedMRtoCTSynthesis2020}
\bibinfo{author}{Yang, H.}, \bibinfo{author}{Sun, J.}, \bibinfo{author}{Carass,
  A.} \& \bibinfo{author}{et~al.}
\newblock \bibinfo{journal}{\bibinfo{title}{Unsupervised {{MR-to-CT Synthesis
  Using Structure-Constrained CycleGAN}}}}.
\newblock {\emph{\JournalTitle{IEEE Transactions on Medical Imaging}}}
  \textbf{\bibinfo{volume}{39}}, \bibinfo{pages}{4249--4261},
  \doiprefix\url{10.1109/TMI.2020.3015379} (\bibinfo{year}{2020}).

\bibitem{liuUnifiedConditionalDisentanglement2021}
\bibinfo{author}{Liu, X.}, \bibinfo{author}{Xing, F.}, \bibinfo{author}{Fakhri,
  G.~E.} \& \bibinfo{author}{et~al.}
\newblock \bibinfo{title}{A {{Unified Conditional Disentanglement Framework}}
  for {{Multimodal Brain MR Image Translation}}}.
\newblock In \emph{\bibinfo{booktitle}{IEEE 18th International Symposium on
  Biomedical Imaging}}, \bibinfo{pages}{10--14},
  \doiprefix\url{10.1109/ISBI48211.2021.9433897} (\bibinfo{year}{2021}).

\bibitem{shenMultiDomainImageCompletion2021}
\bibinfo{author}{Shen, L.}, \bibinfo{author}{Zhu, W.}, \bibinfo{author}{Wang,
  X.} \& \bibinfo{author}{et~al.}
\newblock \bibinfo{journal}{\bibinfo{title}{Multi-{{Domain Image Completion}}
  for {{Random Missing Input Data}}}}.
\newblock {\emph{\JournalTitle{IEEE Transactions on Medical Imaging}}}
  \textbf{\bibinfo{volume}{40}}, \bibinfo{pages}{1113--1122},
  \doiprefix\url{10.1109/TMI.2020.3046444} (\bibinfo{year}{2021}).

\bibitem{hauboldContrastAgentDose2023}
\bibinfo{author}{Haubold, J.}, \bibinfo{author}{Jost, G.},
  \bibinfo{author}{Theysohn, J.~M.} \& \bibinfo{author}{et~al.}
\newblock \bibinfo{journal}{\bibinfo{title}{Contrast {{Agent Dose Reduction}}
  in {{MRI Utilizing}} a {{Generative Adversarial Network}} in an {{Exploratory
  Animal Study}}}}.
\newblock {\emph{\JournalTitle{Investigative Radiology}}}
  \doiprefix\url{10.1097/RLI.0000000000000947} (\bibinfo{year}{2023}).

\bibitem{pasumarthiGenericDeepLearning2021}
\bibinfo{author}{Pasumarthi, S.}, \bibinfo{author}{Tamir, J.~I.},
  \bibinfo{author}{Christensen, S.} \& \bibinfo{author}{et~al.}
\newblock \bibinfo{journal}{\bibinfo{title}{A generic deep learning model for
  reduced gadolinium dose in contrast-enhanced brain {{MRI}}}}.
\newblock {\emph{\JournalTitle{Magnetic Resonance in Medicine}}}
  \textbf{\bibinfo{volume}{86}}, \bibinfo{pages}{1687--1700},
  \doiprefix\url{10.1002/mrm.28808} (\bibinfo{year}{2021}).

\bibitem{schwarzFrequencyBiasGenerative2021a}
\bibinfo{author}{Schwarz, K.}, \bibinfo{author}{Liao, Y.} \&
  \bibinfo{author}{Geiger, A.}
\newblock \bibinfo{title}{On the {{Frequency Bias}} of {{Generative Models}}}.
\newblock In \emph{\bibinfo{booktitle}{Advances in Neural Information
  Processing Systems 34 (NeurIPS 2021)}} (\bibinfo{year}{2021}).

\bibitem{kongBreakingDilemmaMedical2021}
\bibinfo{author}{Kong, L.}, \bibinfo{author}{Lian, C.}, \bibinfo{author}{Huang,
  D.} \& \bibinfo{author}{et~al.}
\newblock \bibinfo{title}{Breaking the {{Dilemma}} of {{Medical Image-to-image
  Translation}}}.
\newblock In \emph{\bibinfo{booktitle}{Advances in Neural Information
  Processing Systems 34 (NeurIPS 2021)}}, vol.~\bibinfo{volume}{34},
  \bibinfo{pages}{1964--1978} (\bibinfo{year}{2021}).

\bibitem{isolaImagetoImageTranslationConditional2018}
\bibinfo{author}{Isola, P.}, \bibinfo{author}{Zhu, J.-Y.},
  \bibinfo{author}{Zhou, T.} \& \bibinfo{author}{et~al.}
\newblock \bibinfo{title}{Image-to-{{Image Translation}} with {{Conditional
  Adversarial Networks}}}.
\newblock In \emph{\bibinfo{booktitle}{IEEE Conference on Computer Vision and
  Pattern Recognition}}, \doiprefix\url{10.1109/CVPR.2017.632}
  (\bibinfo{year}{2017}).

\bibitem{zhuUnpairedImagetoImageTranslation2020}
\bibinfo{author}{Zhu, J.-Y.}, \bibinfo{author}{Park, T.},
  \bibinfo{author}{Isola, P.} \& \bibinfo{author}{et~al.}
\newblock \bibinfo{title}{Unpaired {{Image-to-Image Translation}} using
  {{Cycle-Consistent Adversarial Networks}}}.
\newblock In \emph{\bibinfo{booktitle}{IEEE International Conference on
  Computer Vision}}, \doiprefix\url{10.1109/ICCV.2017.244}
  (\bibinfo{year}{2020}).

\bibitem{yangFDAFourierDomain2020}
\bibinfo{author}{Yang, Y.} \& \bibinfo{author}{Soatto, S.}
\newblock \bibinfo{title}{{{FDA}}: {{Fourier Domain Adaptation}} for {{Semantic
  Segmentation}}}.
\newblock In \emph{\bibinfo{booktitle}{IEEE Computer Society Conference on
  Computer Vision and Pattern Recognition}},
  \doiprefix\url{10.1109/CVPR42600.2020.00414} (\bibinfo{year}{2020}).

\bibitem{jiangFocalFrequencyLoss2021}
\bibinfo{author}{Jiang, L.}, \bibinfo{author}{Dai, B.}, \bibinfo{author}{Wu,
  W.} \& \bibinfo{author}{et~al.}
\newblock \bibinfo{title}{Focal {{Frequency Loss}} for {{Image Reconstruction}}
  and {{Synthesis}}}.
\newblock In \emph{\bibinfo{booktitle}{IEEE International Conference on
  Computer Vision}}, \doiprefix\url{10.1109/ICCV48922.2021.01366}
  (\bibinfo{year}{2021}).

\bibitem{yangFreGANExploitingFrequency2022}
\bibinfo{author}{Yang, M.}, \bibinfo{author}{Wang, Z.}, \bibinfo{author}{Chi,
  Z.} \& \bibinfo{author}{et~al.}
\newblock \bibinfo{journal}{\bibinfo{title}{{{FreGAN}}: {{Exploiting Frequency
  Components}} for {{Training GANs}} under {{Limited Data}}}}.
\newblock {\emph{\JournalTitle{{arXiv}}}}
  \doiprefix\url{10.48550/arXiv.2210.05461} (\bibinfo{year}{2022}).

\bibitem{caiFrequencyDomainImage2021}
\bibinfo{author}{Cai, M.}, \bibinfo{author}{Zhang, H.}, \bibinfo{author}{Huang,
  H.} \& \bibinfo{author}{et~al.}
\newblock \bibinfo{title}{Frequency {{Domain Image Translation}}: {{More
  Photo-realistic}}, {{Better Identity-preserving}}}.
\newblock In \emph{\bibinfo{booktitle}{IEEE International Conference on
  Computer Vision}}, \bibinfo{pages}{13930--13940},
  \doiprefix\url{10.1109/ICCV48922.2021.01367} (\bibinfo{year}{2021}).

\bibitem{maoLeastSquaresGenerative2017}
\bibinfo{author}{Mao, X.}, \bibinfo{author}{Li, Q.}, \bibinfo{author}{Xie, H.}
  \& \bibinfo{author}{et~al.}
\newblock \bibinfo{title}{Least {{Squares Generative Adversarial Networks}}}.
\newblock In \emph{\bibinfo{booktitle}{IEEE International Conference on
  Computer Vision}}, \bibinfo{pages}{2794--2802},
  \doiprefix\url{10.1109/ICCV.2017.304} (\bibinfo{year}{2017}).

\bibitem{balakrishnanVoxelMorphLearningFramework2019}
\bibinfo{author}{Balakrishnan, G.}, \bibinfo{author}{Zhao, A.},
  \bibinfo{author}{Sabuncu, M.~R.} \& \bibinfo{author}{et~al.}
\newblock \bibinfo{journal}{\bibinfo{title}{{{VoxelMorph}}: {{A Learning
  Framework}} for {{Deformable Medical Image Registration}}}}.
\newblock {\emph{\JournalTitle{IEEE Transactions on Medical Imaging}}}
  \textbf{\bibinfo{volume}{38}}, \bibinfo{pages}{1788--1800},
  \doiprefix\url{10.1109/TMI.2019.2897538} (\bibinfo{year}{2019}).

\bibitem{baidRSNAASNRMICCAIBraTS20212021}
\bibinfo{author}{Baid, U.}, \bibinfo{author}{Ghodasara, S.},
  \bibinfo{author}{Mohan, S.} \& \bibinfo{author}{et~al.}
\newblock \bibinfo{journal}{\bibinfo{title}{The {{RSNA-ASNR-MICCAI BraTS}} 2021
  {{Benchmark}} on {{Brain Tumor Segmentation}} and {{Radiogenomic
  Classification}}}}.
\newblock {\emph{\JournalTitle{{arXiv}}}}
  \doiprefix\url{10.48550/arXiv.2107.02314} (\bibinfo{year}{2021}).

\bibitem{menzeMultimodalBrainTumor2015}
\bibinfo{author}{Menze, B.~H.}, \bibinfo{author}{Jakab, A.},
  \bibinfo{author}{Bauer, S.} \& \bibinfo{author}{et~al.}
\newblock \bibinfo{journal}{\bibinfo{title}{The {{Multimodal Brain Tumor Image
  Segmentation Benchmark}} ({{BRATS}})}}.
\newblock {\emph{\JournalTitle{IEEE transactions on medical imaging}}}
  \textbf{\bibinfo{volume}{34}}, \bibinfo{pages}{1993--2024},
  \doiprefix\url{10.1109/TMI.2014.2377694} (\bibinfo{year}{2015}).

\bibitem{bakasAdvancingCancerGenome2017}
\bibinfo{author}{Bakas, S.}, \bibinfo{author}{Akbari, H.},
  \bibinfo{author}{Sotiras, A.} \& \bibinfo{author}{et~al.}
\newblock \bibinfo{journal}{\bibinfo{title}{Advancing {{The Cancer Genome
  Atlas}} glioma {{MRI}} collections with expert segmentation labels and
  radiomic features}}.
\newblock {\emph{\JournalTitle{Scientific Data}}} \textbf{\bibinfo{volume}{4}},
  \bibinfo{pages}{170117}, \doiprefix\url{10.1038/sdata.2017.117}
  (\bibinfo{year}{2017}).

\bibitem{liu3DBrainHeart2022}
\bibinfo{author}{Liu, Y.}, \bibinfo{author}{Dwivedi, G.},
  \bibinfo{author}{Boussaid, F.} \& \bibinfo{author}{et~al.}
\newblock \bibinfo{journal}{\bibinfo{title}{{{3D Brain}} and {{Heart Volume
  Generative Models}}: {{A Survey}}}}.
\newblock {\emph{\JournalTitle{{arXiv}}}}
  \doiprefix\url{10.48550/arXiv.2210.05952} (\bibinfo{year}{2022}).

\bibitem{xieCrossModalityNeuroimageSynthesis2022}
\bibinfo{author}{Xie, G.}, \bibinfo{author}{Wang, J.}, \bibinfo{author}{Huang,
  Y.} \& \bibinfo{author}{et~al.}
\newblock \bibinfo{journal}{\bibinfo{title}{Cross-{{Modality Neuroimage
  Synthesis}}: {{A Survey}}}}.
\newblock {\emph{\JournalTitle{{arXiv}}}}
  \doiprefix\url{10.48550/arXiv.2202.06997} (\bibinfo{year}{2022}).

\bibitem{aliRoleGenerativeAdversarial2022}
\bibinfo{author}{Ali, H.}, \bibinfo{author}{Biswas, M.~R.},
  \bibinfo{author}{Mohsen, F.} \& \bibinfo{author}{et~al.}
\newblock \bibinfo{journal}{\bibinfo{title}{The role of generative adversarial
  networks in brain {{MRI}}: A scoping review}}.
\newblock {\emph{\JournalTitle{Insights into Imaging}}}
  \textbf{\bibinfo{volume}{13}}, \bibinfo{pages}{98},
  \doiprefix\url{10.1186/s13244-022-01237-0} (\bibinfo{year}{2022}).

\bibitem{detlefsenTorchMetricsMeasuringReproducibility2022}
\bibinfo{author}{Detlefsen, N.~S.}, \bibinfo{author}{Borovec, J.},
  \bibinfo{author}{Schock, J.} \& \bibinfo{author}{et~al.}
\newblock \bibinfo{journal}{\bibinfo{title}{{{TorchMetrics}} - {{Measuring
  Reproducibility}} in {{PyTorch}}}}.
\newblock {\emph{\JournalTitle{Journal of Open Source Software}}}
  \textbf{\bibinfo{volume}{7}}, \bibinfo{pages}{4101},
  \doiprefix\url{10.21105/joss.04101} (\bibinfo{year}{2022}).

\bibitem{kingmaAdamMethodStochastic2017}
\bibinfo{author}{Kingma, D.~P.} \& \bibinfo{author}{Ba, J.}
\newblock \bibinfo{journal}{\bibinfo{title}{Adam: {{A Method}} for {{Stochastic
  Optimization}}}}.
\newblock {\emph{\JournalTitle{{arXiv}}}}
  \doiprefix\url{10.48550/arXiv.1412.6980} (\bibinfo{year}{2017}).

\bibitem{paszkePyTorchImperativeStyle2019}
\bibinfo{author}{Paszke, A.}, \bibinfo{author}{Gross, S.},
  \bibinfo{author}{Massa, F.} \& \bibinfo{author}{et~al.}
\newblock \bibinfo{title}{{{PyTorch}}: {{An Imperative Style}},
  {{High-Performance Deep Learning Library}}}.
\newblock In \emph{\bibinfo{booktitle}{Advances in Neural Information
  Processing Systems 32 (NeurIPS 2019)}} (\bibinfo{year}{2019}).

\bibitem{Falcon_PyTorch_Lightning_2019}
\bibinfo{author}{Falcon, W.} \& \bibinfo{author}{{the PyTorch Lightning team}}.
\newblock \bibinfo{title}{Pytorch lightning},
  \doiprefix\url{10.5281/zenodo.7469930} (\bibinfo{year}{2019}).

\end{thebibliography}
\section*{Author contributions statement}
I.M.B., M.L., F.K., and A.H. conceived the method and experiments, I.M.B. and A.H. conducted the experiments, I.M.B., M.L., M.D., A.H., analyzed the results. All authors reviewed the manuscript.
\section*{Additional information}
\subsection*{Competing interests}
All authors are employees of Bayer AG, Berlin, Germany.
\subsection*{Data Availability}
The dataset (i.e., BraTS2021) analyzed during the current study are available from http://braintumorsegmentation.org/.
\end{document}